\documentstyle[11pt]{article}
\newcommand{\qed}{\ \ \ \rule{7pt}{8pt}\medskip}

\newcommand{\comment}[1]{}

\newcommand{\real}{\ifmmode {\rm R} \else ${\rm R}$ \fi}
\newcommand{\nat}{\ifmmode {\rm N} \else ${\rm N}$  \fi}
\newcommand{\tot}{\ifmmode {\cal T} \else ${\cal T}$ \fi}
\newcommand{\sigstar}{\ifmmode \Sigma^{\ast} \else $\Sigma^{\ast}$ \fi}

\newtheorem{theorem}{Theorem}
\newtheorem{lemma}[theorem]{Lemma}

\newtheorem{claim}[theorem]{Claim}

\newlength{\thislabel}

\def\lablimer2stlabel#1{\rm #1\hfil}

\title{On The Closest String and Substring Problems
\footnote{Some of the results
in this paper have been
presented in {\em Proc.\ 31st ACM Symp.\ Theory of Computing}, May,
1999 \cite{LMW99},
and in {\em Proc.\ 11th Symp. Combinatorial Pattern Matching}, June,
2000, \cite{M00}.}}
\author{
Ming Li \\
Department of Computer Science\\
University of Waterloo\\
Waterloo, Ont. N2L 3G1, Canada\\
 E-mail: mli@math.uwaterloo.ca
\and
Bin Ma\\
Department of Computer Science \\
University of Waterloo\\
Waterloo, Ont. N2L 3G1, Canada\\
E-mail: b3ma@wh.math.uwaterloo.ca
\and
Lusheng Wang\\
Department of Computer Science \\
City University of Hong Kong \\
Kowloon, Hong Kong \\
E-mail: lwang@cs.cityu.edu.hk
}
\date{}

\begin{document}
\maketitle

\begin{abstract}
The problem of finding a center string that is `close' to every
given string arises and has many applications in computational molecular
biology and coding theory.

This problem has two versions: the Closest String
problem and the Closest Substring problem.
Assume that we are given a set of strings
${\cal S}=\{s_1, s_2, \ldots, s_n\}$ of strings, say, each of length $m$.
The Closest String problem~\cite{BLPR97,BGHMS97,FL97,GJL99,LL+99}
asks for the smallest $d$ and a string $s$  of length $m$ which is within
Hamming distance $d$ to each $s_i\in {\cal S}$.
This problem comes from coding theory when we are looking for a code
not too far away from a given set of codes \cite{FL97}.
The problem is NP-hard~\cite{FL97,LL+99}. Berman {\em et al}
\cite{BGHMS97} give a polynomial time
algorithm for constant $d$. For super-logarithmic $d$,
Ben-Dor {\em et al} \cite{BLPR97} give an efficient approximation algorithm
using linear program relaxation technique.
The best polynomial time approximation has ratio $\frac{4}{3}$
for all $d$, given by \cite{LL+99} and \cite{GJL99}.
The Closest Substring problem looks for a string $t$ which is
within Hamming distance $d$ away from a substring of each $s_i$.
This problem only has a $2- \frac{2}{2|\Sigma|+1}$ approximation
algorithm previously \cite{LL+99} and
is much more elusive than the Closest String problem, but
it has many applications in finding
conserved regions, genetic drug target identification, and genetic probes
in molecular biology~\cite{HS94,LR90,LB+91,PBPR89,PH96,
S90,SH91,W86,WAG84,WP84,LL+99}. Whether there are efficient
approximation algorithms for both problems are
major open questions in this area.

We present two polynomial time approxmation algorithms with
approximation ratio $1+ \epsilon$ for any small $\epsilon$
to settle both questions.

\end{abstract}

\section{Introduction}
\label{sec-intro}
Many problems in molecular biology involve finding similar
regions common to each sequence in a given set
of DNA, RNA, or protein sequences. These problems find applications in
locating binding sites and finding conserved
regions in unaligned sequences~\cite{SH91,LR90,HS94,S90},
genetic drug target identification~\cite{LL+99}, designing genetic
probes \cite{LL+99}, universal PCR primer design~\cite{LB+91,DR+93,PH96,LL+99},
and, outside computational biology, in coding theory~\cite{FL97,GJL99}.
Such problems may be considered to be various generalizations of the common
substring problem, allowing errors.
Many objective functions have been proposed
for finding such regions common to every given strings.
A popular and most fundamental measure is the Hamming distance. Other
measures, like the relative entropy measure used
by Stormo and his coauthors \cite{HS94} may be
considered as generalizations of Hamming distance, requires
different techniques, and is considered in \cite{LMW99-j}.

Let $s$ and $s'$ be finite strings.
Let $d(s,s')$ denote the Hamming distance between $s$ and $s'$.
$|s|$ is the length of $s$. $s[i]$ is the $i$-th character
of $s$. Thus, $s=s[1]s[2] \ldots s[|s|]$.
The following are the problems we study in this paper:

\vspace{1ex}
\noindent
{\sc Closest String:} Given a set ${\cal S}=\{s_1,s_2,\ldots,s_n\}$
of strings each of length $m$, find a
center string $s$ of length $m$ minimizing $d$
such that for every string $s_{i}\in {\cal S}$, $d(s, s_{i})\leq d$.

\vspace{1ex}
\noindent
{\sc Closest Substring:} Given a set ${\cal S}=\{s_1,s_2,\ldots,s_n\}$
of strings, and an integer $L$, find a center string $s$ of length $L$
minimizing $d$ such that for each $s_i\in {\cal S}$ there is
a length $L$ substring $t_i$ of $s_i$ with  $d(s, t_{i})\leq d$.
\vspace{1ex}

{\sc Closest String} has been widely and independently studied
in different contexts. In the context of coding theory
it was shown to be NP-hard~\cite{FL97}. In DNA sequence related topics,
\cite{BGHMS97} gave an exact algorithm when the distance $d$ is a constant.
\cite{BLPR97,GJL99} gave near-optimal
approximation algorithms only for large $d$ (super-logarithmic in number of
sequences); however the straightforward linear programming relaxation technique
does not work when $d$ is small because the randomized
rounding procedure introduces large errors.
This is exactly the reason why \cite{GJL99,LL+99}
analyzed more involved approximation algorithms, and
obtained the ratio $\frac{4}{3}$ approximation algorithms.
Note that the small $d$ is key in applications such as
genetic drug target search where we look for similar regions to which
a complementary drug sequence would bind. It is a major open
problem~\cite{FL97,BGHMS97,BLPR97,GJL99,LL+99} to achieve the best
approximation ratio for this problem. (Justifications for using Hamming
distance can also be found in these references, especially \cite{LL+99}.)
We present a polynomial approximation scheme (PTAS), settling the problem.

{\sc Closest Substring} is a more general version of the
{\sc Closest String} problem. Obviously, it is also NP-hard.
In applications such as drug target identification and
genetic probes design, the radius $d$ is usually small.
Moreover, when the radius $d$ is small,
the center strings can also be used
as {\it motifs} in {\it repeated-motif} methods  for
multiple sequence alignment problems
~\cite{Danbook, PBPR89,SAL91,W86,WAG84,WP84},
that repeatedly find motifs and recursively decompose the sequences
into shorter sequences.
A trivial ratio-$2$ approximation was given in~\cite{LL+99}.
We presented the first nontrivial
algorithm with approximation ratio $2- \frac{2}{2|\Sigma | +1}$,
in \cite{LMW99}. This is a key open problem in search of a
potential genetic drug sequence which is ``close''
to some sequences (of harmful germs) and ``far'' from some other sequences
(of humans). The problem appears to be much more elusive than
{\sc Closest String}.
We extend the techniques developed  for closest string here
to design a PTAS for closest substring problem  when
$d$ is small, i.e., $d\leq O(\log N)$, where $N$
is the input size of the instance.
Using a {\em random sampling} technique, and combining our
methods for {\sc Closest String}, we then design a PTAS
for {\sc Closest Substring}, for all $d$.

\section{Approximating {\sc Closest String}}
\label{sec-closeststring}
In this section, we give a PTAS for {\sc Closest String}.
We note that a direct application of LP relaxation
in \cite{BLPR97} does not work when the optimal solution is small.
Rather we extend an idea in \cite{LL+99} to do LP relaxation
only to a fraction of the bits.
Let ${\cal S}=\{s_1, s_2, \ldots, s_n\}$ be a set of  $n$ strings each of
length $m$.

The idea is as follows. Let $r$ be a constant.
If we choose a subset
of $r$ strings from ${\cal S}$, consider the bits that they all agree.
Intutively, we can replace the corresponding bits
in the optimal solution by these bits of the $r$ strings,
and this will only slightly worsen the solution.
Lemma~\ref{KEY} shows that this is true for at least one subset
of $r$ strings. Then all we
need to do is to optimize on the positions (bits) where they do not agree,
by LP relaxation and randomized rounding.

We first introduce some notations.  Let
$P=\{j_1,j_2,\ldots,j_k\}$ be a set (multiset) and
$1\leq j_1 \leq j_2 \leq \cdots \leq j_k\leq m$.
$P$ is called a {\it position set} ({\it multiset}).
Let $s$ be a string of length $m$,
then $s|_P$ is the string $s[j_1]\,s[j_2]\,\cdots \,s[j_k]$.

For any  $k \geq 2$, let $1 \leq i_1,i_2,\ldots,i_k \leq n$ be
$k$ distinct numbers.
Let $Q_{i_1,i_2,\ldots,i_k}$ be the set of positions
where $s_{i_1},s_{i_2},\ldots,s_{i_k}$ agree.
Obviously $|Q_{i_1,i_2,\ldots,i_k}| \geq m- kd_{opt}$.
Let $\rho_0 = \max _{1\leq i,j\leq n} {d(s_i,s_j)}/{d_{opt}}$.
The following lemma is the key of our approximation algorithm.
\begin{lemma}\label{KEY}
If $\rho_0>  1+\frac{1}{2r-1}$, then
for any constant $r$, there are indices
$1 \leq i_1,i_2,\ldots,i_r \leq n$ such that for any $1 \leq l \leq n$,
$$
d(s_l|_{Q_{i_1,i_2,\ldots,i_r}}, s_{i_1}|_{Q_{i_1,i_2,\ldots,i_r}})
- d(s_l|_{Q_{i_1,i_2,\ldots,i_r}}, s|_{Q_{i_1,i_2,\ldots,i_r}})
\leq \frac 1{2r-1} d_{opt}.
$$
\end{lemma}
\begin{proof}
Let $p_{i_1,i_2,\ldots,i_k}$ be the number of mismatches between
$s_{i_1}$ and $s$ at the positions in $Q_{i_1,i_2,\ldots,i_k}$.  Let
$
\rho _k = \min _{1\leq i_1,i_2,\ldots,i_k\leq n}
 {p_{i_1,i_2,\ldots,i_k}} /{d_{opt}}.
$
First, we prove the following claim.
\begin{claim}\label{Fact1}
For any $k$ such that $2 \leq k \leq r$, where $r$ is the constant
in the algorithm closestString, there are indices
$1 \leq i_1,i_2,\ldots,i_r \leq m$ such that for any $1 \leq l \leq n$.
$$
 | \{j \in Q_{i_1,i_2,\ldots,i_r} \,|\,
 s_{i_1}[j] \neq s_l[j] \mbox{ and } s_{i_1}[j]
\neq s[j]\} |
\leq  (\rho _k - \rho _{k+1}) \,d_{opt}
$$
\end{claim}
\begin{proof}
Consider indices $1\leq i_1,i_2,\ldots,i_k\leq m$ such that
${p_{i_1,i_2,\ldots, i_k}}= \rho _k d_{opt}$.
Then for any $1\leq i_{k+1},i_{k+2},\ldots,i_{r} \leq m$
and $1\leq l\leq n$, we have
\begin{eqnarray}
&& | \{j \in Q_{i_1,i_2,\ldots,i_r} \,|\,
 s_{i_1}[j] \neq s_l[j] \mbox{ and } s_{i_1}[j] \neq s[j]\}
|
\nonumber\\
&\leq & | \{j \in Q_{i_1,i_2,\ldots,i_k} \,|\,
 s_{i_1}[j] \neq s_l[j] \mbox{ and }  s_{i_1}[j] \neq s[j]\} | \label{eq-tmp11}\\
&=&
| \{ j \in Q_{i_1,i_2,\ldots,i_k} \,|\, s_{i_1}[j] \neq s[j] \}
 - \{ j \in Q_{i_1,i_2,\ldots,i_k} \,|\,
s_{i_1}[j]=s_l[j] \mbox{ and } s_{i_1}[j] \neq s[j] \} |
\nonumber\\
&=&
| \{ j \in Q_{i_1,i_2,\ldots,i_k} \,|\, s_{i_1}[j] \neq s[j] \}
 - \{ j \in Q_{i_1,i_2,\ldots,i_k,l} \,|\, s_{i_1}[j] \neq s[j] \}|
\nonumber\\
&=&
p_{i_1,i_2,\ldots,i_k} - p_{i_1,i_2,\ldots,i_k,l}
\label{eq-tmp12}\\
&\leq& (\rho _k - \rho _{k+1}) \,d_{opt},
\nonumber
\end{eqnarray}
where Inequality (\ref{eq-tmp11}) is  from the fact that
$Q_{i_1,i_2,\ldots,i_r} \subseteq Q_{i_1,i_2,\ldots,i_k}$
and Equality (\ref{eq-tmp12}) is  from  the fact that
$Q_{i_1,i_2,\ldots,i_k,l} \subseteq Q_{i_1,i_2,\ldots,i_k}$.
$\Box$
\end{proof}

\begin{claim}\label{Fact2}
$\min \{\rho_0 -1, \rho _2 - \rho_3, \rho _3 - \rho_4,
\ldots , \rho _r - \rho_{r+1} \} \leq \frac {1}{2 r-1}.$
\end{claim}
\begin{proof}
Consider $1\leq i,j \leq n$ such that
$d(s_i,s_j) = \rho_0 d_{opt}$.  Then among the
positions where $s_i$ mismatches $s_j$, for at least one of the
two strings, say, $s_i$, the number of
mismatches between $s_i$ and $s$ is at least
$\rho_0 d_{opt}/2$.  Thus, among the positions where
$s_i$ matches $s_j$, the number of mismatches between
$s_i$ and $s$ is at most $(1-\frac{\rho_0}{2}){d_{opt}}$.
Therefore, $\rho_2 \leq 1-\frac {\rho_0}{2}$. So,
$$
\frac {{\frac 12} (\rho_0 -1) + ( \rho _2 - \rho_3) + ( \rho _3 - \rho_4)
   + \cdots + (\rho_r - \rho_{r+1}) }{\frac 12 +r-1}
\leq \frac {{\frac 12} \rho _0 + \rho _2 - \frac 12 }{r-\frac 12}
\leq  \frac {1}{2 r-1}
$$
Thus, at least one of $\rho_0-1$, $\rho _2 - \rho_3$, $\rho _3 - \rho_4$,
$\ldots$, $\rho _r - \rho_{r+1}$ is
less than or equal to
$\frac {1}{2 r-1}$.
$\Box$
\end{proof}

If $\rho_0 > 1+\frac {1}{2 r -1}$,
them from Claim    \ref{Fact2},
 there must be a $2 \leq k \leq r$ such that
$\rho _k - \rho _{k+1} \leq \frac {1}{2 r -1}$.
 From Claim \ref{Fact1},
$$
 | \{j \in Q_{i_1,i_2,\ldots,i_r} \,|\,
 s_{i_1}[j] \neq s_l[j] \mbox{ and } s_{i_1}[j]
\neq s[j]\} |
\leq  \frac{1}{2r-1} \,d_{opt} \ .
$$
Hence, there are at most $\frac 1{2r-1} \, d_{opt}$ bits
in $Q_{i_1,i_2,\ldots,i_r}$
where $s_l$ differs from $s_{i_1}$ while agrees with
$s$.  The lemma is proved.
$\qed$
\end{proof}

Lemma \ref{KEY} hints us  to select $r$ strings
$s_{i-1}, s_{i_2}, \ldots, s_{i_r}$  from $\cal {S}$
at a time and  use the unique letters at the positions in
 $Q_{i_1,i_2,\ldots, i_r}$  as an approximation of
the optimal center string $s$.
For the  positions in $P_{i_1,i_2,\ldots, i_r}=\{1,2,\ldots, L\}-Q_{i_1,i_2,\ldots, i_r}$,
we use ideas in \cite{LL+99}, i.e., the following   two strategies:
(1) if $|P_{i_1,i_2,\ldots, i_r}|$ is small, i.e., $d\leq O(\log L)$,
we can enumerate $|\Sigma| ^{|P_{i_1,i_2,\ldots, i_r}|}$ possibilities  to approximate $s$;
(2) if $|P_{i_1,i_2,\ldots, i_r}|$ is large, i.e., $d>O(\log L)$, we use the
LP relaxation to approximate $s$.  The details are
found in Lemma~\ref{lem-rest}.
Before
presenting our main result, we need the following
two lemmas, where Lemma~\ref{lem-chernoff} is commonly known
as Chernoff's bounds~(\cite{MR95}, Theorem~4.2 and 4.3):
\begin{lemma}
{\rm \cite{MR95}~}
Let $X_1,X_2,\ldots,X_n$ be $n$ independent random 0-1 variables,
where $X_i$ takes $1$ with probability $p_i$, $0<p_i<1$.
Let $X=\sum _{i=1}^n X_i$, and $\mu=E[X]$.
Then for any $\delta>0$,
\begin{enumerate}
\item[(1)]
${\bf Pr}(X>(1+\delta) \mu ) < \left[\frac {{\bf e}^{\delta}} {(1+\delta)^{(1+ \delta)}}\right]^{\mu}$,
\item[(2)]
${\bf Pr}(X<(1-\delta) \mu ) \leq \exp \left( -\frac 12 \mu \delta ^2 \right)$.
\vspace{-6pt}
\end{enumerate}
\label{lem-chernoff}
\end{lemma}

 From Lemma~\ref{lem-chernoff}, we can prove the following lemma:
\begin{lemma}
Let $X_i$, $X$ and $\mu$ be defined as in Lemma~\ref{lem-chernoff}.
Then for any $0<\epsilon\leq 1$,
\begin{enumerate}
\item[(1)]
${\bf Pr}(X>\mu+\epsilon\, n ) < \exp \left(-\frac 13 n \epsilon ^2 \right)$,
\item[(2)]
${\bf Pr}(X<\mu-\epsilon\,n ) \leq \exp \left( -\frac 12 n \epsilon ^2 \right)$.
\vspace{-6pt}
\end{enumerate}
\label{lem-chernoff1}
\end{lemma}
\begin{proof}
(1) Let $\delta = \frac {\epsilon n} {\mu}$.  By Lemma~\ref{lem-chernoff},
$$
{\bf Pr}(X>\mu +\epsilon n )
<\left[\frac {{\bf e}^{\frac{\epsilon n}{\mu}}}
{(1+\frac{\epsilon n}{\mu})^{(1+\frac{\epsilon n}{\mu})}}\right]^{\mu}
=\left[\frac {\bf e}{(1+\frac{\epsilon n}{\mu})^{(1+\frac{\mu}{\epsilon n})}}\right]^{\epsilon n}
\leq \left[\frac {\bf e} {(1+\epsilon)^{1+\frac 1{\epsilon}}} \right]^{\epsilon n},
$$
where the last inequality is because $\mu \leq n$ and
that $(1+x)^{(1+\frac 1x)}$ is increasing for $x \geq 0$.
It is easy to verify that for $0< \epsilon \leq 1$,
$\frac {\bf e} {(1+\epsilon)^{1+\frac 1{\epsilon}}}
\leq \exp \left(-\frac \epsilon 3 \right).$
Therefore, (1) is proved.

(2)  Let $\delta =\frac {\epsilon n} {\mu}$.  By Lemma~\ref{lem-chernoff},
(2) is proved.
$\qed$
\end{proof}

Now, we come back to the approximation of $s$ at the positions
in  $P_{i_1,i_2,\ldots,i_r}$.

\begin{lemma}
\label{lem-rest}
Let  ${\cal S} = \{s_1, s_2, \ldots s_n\}$, where $|s_i| = m$ for all $i$.
Assume that $s$ is the optimal solution of {\sc Closest String}
and $\max_{1 \leq i \leq n} d(s_i,s) =d_{opt}$.
Given  a  string $s'$ and a  position set $Q$ of size $m-O(d_{opt})$
such that for any $i=1, \ldots , n$
\begin{equation}
\label{eq-rest01}
d(s_i|_Q,s'|_Q)-d(s_i|_Q,s|_Q) \leq \rho \, d_{opt},
\end{equation}
where  $0 \leq \rho \leq 1$,
one can obtain a solution with cost at most
$(1+\rho + \epsilon)d_{opt}$ in polynomial time
 for any fixed $\epsilon \geq 0$.
\end{lemma}

\begin{proof}
Let $P=\{1,2,\ldots,m\} - Q$.  Then, for any two
strings $x$ and  $x'$ of length $m$, we have
$d(x|_P,x'|_P)+d(x|_Q,x'|_Q)=d(x,x')$.
Thus for any $i=1,2, \ldots ,n$,
$$
d(s_i|_P,s|_P)=d(s_i,s)-d(s_i|_Q,s|_Q)
\leq (1+ \rho)\,d_{opt} - d(s_i|_Q,s'|_Q).
$$
Therefore,  the  following optimization problem
\begin{equation}
\label{lps1}
\left\{
 \begin{array}{l}
 \min \;\; d;\\
 d(s_i|_P,x) \leq d - d(s_i|_Q,s'|_Q), \;\;
  i=1, \cdots , n; |x|=|P |,
 \end{array}
\right.
\end{equation}
 has a solution with cost
$d \leq (1+ \rho) d_{opt}$.
Suppose that the optimization problem  has an optimal solution $x$ such that
$d=d_0$.  Then
\begin{equation}
\label{eq-d0}
d_0\leq (1+\rho) d_{opt}.
\end{equation}
Now we solve (\ref{lps1}) approximately.
Similar to \cite{BLPR97,LL+99}, we use a 0-1 variable
$x_{j,a}$ to indicate whether $x[j]=a$.  Denote
$\chi (s_i[j],a)=0$ if $s_i[j]=a$ and $1$ if $s_i[j] \neq a$.
Then (\ref{lps1}) can be rewritten as a 0-1 optimization problem
as follows:
\begin{equation}
\label{lps2}
\left\{
 \begin{array}{l}
 \min \;\; d;\\
 \sum _{a \in \Sigma} x_{j,a}=1, \;\; j=1,2,\ldots,|P|,\\
 \sum_{1\leq j\leq |P|} \sum _{a \in \Sigma}
   \chi (s_i[j],a) \, x_{j,a}
   \leq d - d(s_i|_Q,s'|_Q), \;\; i=1,2, \ldots , n.
 \end{array}
\right.
\end{equation}
Solve (\ref{lps2}) by linear programming to
get a fractional solution ${\bar x}_{j,a}$ with cost
${\bar d}$.  Clearly ${\bar d} \leq d_0$.
Independently for each $0 \leq j \leq |P|$,
with probability ${\bar x}_{j,a}$,
set $x_{j,a}=1$ and $x_{j,a'}=0$ for any $a' \neq a$.
Then we get a solution $x_{j,a}$ for the 0-1 optimization
problem, hence a solution $x$ for (\ref{lps1}).
It is easy to see that
$\sum _{a \in \Sigma} \chi (s_i[j],a) \, x_{j,a} $
takes $1$ or $0$ randomly and independently for
different $j$'s.  Thus
$d(s_i|_P,x)= \sum_{1\leq j\leq |P|} \sum _{a \in \Sigma}
   \chi (s_i[j],a) \, x_{j,a}$
is a sum of $|P|$ independent 0-1 random variables, and
\begin{eqnarray}
E[d(s_i|_P,x)]&=&\sum_{1\leq j\leq |P|} \sum _{a \in \Sigma}
   \chi (s_i[j],a) \, E[x_{j,a}]
\nonumber\\
&=&\sum_{1\leq j\leq |P|} \sum _{a \in \Sigma}
   \chi (s_i[j],a) \, {\bar x}_{j,a}
\nonumber\\
&\leq& {\bar d}-d(s_i|_Q,s'|_Q)\leq d_0-d(s_i|_Q,s'|_Q).
\label{eq-hoeff01}
\end{eqnarray}
Therefore, for any fixed $\epsilon '>0$,
by Lemma~\ref{lem-chernoff1},
$$
{\bf Pr}\left(
d(s_i|_P,x) \geq d_0+\epsilon' |P|
-d(s_i|_Q,s'|_Q)
        \right)
\leq
\exp \left(- \frac 13 {{\epsilon '}^2} |P|\right).
$$
Considering all sequences, we have
\[
{\bf Pr}\left(d(s_i|_P,x) \geq d_0 +\epsilon ' |P|-
d(s_i|_Q,s'|_Q)
     ~ {\rm for~at~least~one~}i       \right)
\leq n\times \exp \left(-\frac 13 {\epsilon'} ^2 |P|\right).
\]
If $|P| \geq (4 \ln n )/ {\epsilon'} ^2$,
then,
$n\times
 \exp\left(-\frac 13 {\epsilon'} ^2 |P|\right) \leq n^{-\frac 13}$.
Thus
we obtain a randomized algorithm to find a solution
for (\ref{lps1}) with cost at most
$d_0 +\epsilon' |P|$ with probability at least $1-n^{-\frac
13}$.
The above randomized algorithm can be derandomized
by standard method of conditional probabilities~\cite{MR95}.

If $|P| < ( 4 \ln n )/ {\epsilon'} ^2$,
$|\Sigma| ^{|P|} < n ^{(4 \ln |\Sigma|)/{\epsilon'} ^2}$
is a polynomial of $n$.
So, we can enumerate all strings in $\Sigma ^{|P|}$ to find
an optimal solution for (\ref{lps1}).
Thus, in both cases, we can obtain a solution $x$ for the optimization
problem (\ref{lps1}) with cost at most
$d_0+\epsilon' |P|$ in polynomial time.
Since $|P|=O(d_{opt})$,
$|P| \leq c\times d_{opt}$ for a {\em constant} $c$.
Let $\epsilon'= \frac {\epsilon}{c}$
and
$s^*=R(s',x,P)$. From Formula (\ref{lps1}),
\begin{eqnarray}
d(s_i,s^*)
&=&d(s_i|_P,s^*|_P)+d(s_i|_Q,s^*|_Q)
\nonumber\\
&=&d(s_i|_P,x)+d(s_i|_Q,s'|_Q)
\nonumber\\
&\leq& d_0+ \epsilon' |P|
\leq (1+ \rho) d_{opt} + \epsilon d_{opt},
\nonumber
\end{eqnarray}
where the last inequality is from Formula~(\ref{eq-d0}).
This proves the lemma.
$\Box$
\end{proof}

Now we describe the complete algorithm in Figure~\ref{stringAlg}.

\begin{figure}[ht]
\begin{center}
\begin{tabular}{|l|}
\hline
\multicolumn{1}{|c|}{\bf Algorithm ~closestString}
\\
\makebox[.45in][l]{{Input}} \parbox[t]{4.55in}
{$s_1, s_2, \ldots , s_n \in \Sigma^m$.}
\\
\makebox[.45in][l]{{Output}} \parbox[t]{4.55in}
{a center string $s \in \Sigma^m$.}
\\
\makebox[.2in][l]{1.} \parbox[t]{4.8in}
{{\bf for} each $r$-element subset $\{ s_{i_1}$, $s_{i_2}$, $\ldots$,
$s_{i_r} \}$ of the $n$ input strings {\bf do}}
\\
\makebox[.5in][r]{(a)} \parbox[t]{4.5in}
{$Q=\{1\leq j \leq m \,|\, s_{i_1}[j]=s_{i_2}[j]=\ldots = s_{i_r}[j] \}$,
$P=\{1,2,\ldots, m\} - Q$.}
\\
\makebox[.5in][r]{(b)} \parbox[t]{4.5in}
{Solve the optimization problem defined by Formula (\ref{lps1})
as described in the proof of Lemma~\ref{lem-rest} to get an approximate
solution $x$ of length $|P|$.}
\\
\makebox[.5in][r]{(c)} \parbox[t]{4.5in}
{Let $s'$ be a string such that $s'|_Q=s_{i_1}|_Q$ and $s'|_P=x$.
Calculate the cost of $s'$ as the center
string.}
\\
\makebox[.2in][l]{2.} \parbox[t]{4.8in}
{{\bf for}   $i=1, 2, \ldots , n$ {\bf do}}
\\
\makebox[.4in][l]{}\parbox[t]{4.6in}
{calculate the cost of $s_i$ as the center string.}
\\
\makebox[.2in][l]{3.} \parbox[t]{4.8in}
{Output the best solution of the above two steps.}
\\
\hline
\end{tabular}
\caption{Algorithm for {\sc Closest String}}
\label{stringAlg}
\end{center}
\end{figure}

\begin{theorem}
\label{th-uniform}
The algorithm closestString is a PTAS for {\sc Closest String}.
\end{theorem}
\begin{proof}
Given an instance of {\sc Closest String},
suppose $s$ is an optimal solution and the optimal
cost is $d_{opt}$, i.e. $d(s,s_i) \leq d_{opt}$ for all $i$.
Let $P$ be defined as step 1(a) of Algorithm~closestString.
Since for every position in $P$, at least one of the $r$
strings $s_{i_1},s_{i_2},\ldots,s_{i_r}$ conflict
the optimal center string $s$, so we have
$|P|\leq r\times d_{opt}$.  As far as $r$ is a constant,
step 1(b) can be done in polynomial time by Lemma~\ref{lem-rest}.
Obviously the other steps of
Algorithm~closestString runs in polynomial time,
with $r$ as a constant.

If $\rho_0 -1 \leq \frac {1}{2 r -1}$,
then by the definition of $\rho _0$, it is easy to see that
the algorithm finds a solution with cost at most
$\rho_0 d_{opt} \leq (1+ \frac {1}{2r-1}) d_{opt}$ in step 2.

If $\rho_0 > 1+\frac {1}{2 r -1}$,
them from  Lemma \ref{KEY} and Lemma \ref{lem-rest},
the algorithm finds a solution with cost at most
$(1+\frac 1{2r-1} + \epsilon)d_{opt}$. This proves the theorem.
$\Box$
\end{proof}

\section{Approximating {\sc Closest Substring}  when $d$ is small}
In some applications such as drug target identification,
genetic probe design, the radius $d$ is often small.
As a direct application of Lemma \ref{KEY},
we now present a PTAS for {\sc Closest String}
when the radius $d$ is small, i.e., $d<O(\log N)$, where $N$
stands for the input size of the instance.
Again, we focus on the construction of the center string.
The basic idea is  to choose $r$ substrings
$t_{i_1}$, $t_{i_2}$, $\ldots$, $t_{i_r}$
of length $L$
from the strings in ${\cal S}$,
keep the letters at the positions where
$t_{i_1}$, $t_{i_2}$, $\ldots$, $t_{i_r}$ all agree, and
try all possibilities for the rest of the positions.
The complete algorithm is described in Figure~\ref{fig-Algsmall}:

\begin{figure}[h]
\begin{center}
\begin{tabular}{|l|}
\hline
\multicolumn{1}{|c|}{\bf Algorithm ~smallSubstring}
\\
\makebox[.45in][l]{{Input}} \parbox[t]{4.55in}
{$s_1, s_2, \ldots , s_n \in \Sigma^m$.}
\\
\makebox[.45in][l]{{Output}} \parbox[t]{4.55in}
{a center string $s \in \Sigma^L$.}
\\
\makebox[.2in][l]{1.} \parbox[t]{4.8in}
{{\bf for} each $r$-element subset $\{t_{i_1}$, $t_{i_2}$, $\ldots$,
$t_{i_r} \}$, where $t_{i_j}$ is a substring of length $L$  from
$s_{i_j}$ {\bf do}}
\\
\makebox[.5in][r]{(a)} \parbox[t]{4.5in}
{$Q=\{1\leq j \leq m \,|\, t_{i_1}[j]=t_{i_2}[j]=\ldots = t_{i_r}[j] \}$,
$P=\{1,2,\ldots, m\} - Q$.}
\\

\makebox[.5in][r]{(b)} \parbox[t]{4.5in}
{{\bf for} every $x \in \Sigma ^{|P|}$ {\bf do}}
\\
\makebox[.7in][r]{} \parbox[t]{4.3in}
{let $t=S(t_{i_1},x,P)$; compute the cost of the solution $t$.}
\\
\makebox[.2in][l]{2.} \parbox[t]{4.8in}
{{\bf for} every length $L$ substring $t_k$ from any given sequence {\bf do}}
\\
\makebox[.4in][r]{} \parbox[t]{4.6in}
{compute the cost of the solution with $t_k$ as the center string}\\
\makebox[.2in][l]{3.} \parbox[t]{4.8in}
{select a center string  that leads  the best result in Step 1 and
Step 2;
output the best solution of the above two steps.}
\\
\hline
\end{tabular}
\caption{Algorithm for {\sc Closest Substring} when $d$ is
small}
\label{fig-Algsmall}
\end{center}
\end{figure}

\begin{theorem}
Algorithm smallSubstring is a PTAS for {\sc Closest Substring}
when the radius $d$ is small, i.e., $d\leq O(\log N)$, where $N$
is the input size.
\end{theorem}
\begin{proof}
Obviously, the size of $P$ in Step 1 is at most $O(r\times \log N)$.
Step 1 takes $O((mn)^{r}\times\Sigma ^{O(r \times \log N)}\times mnL)
=O(N^{r+1} \times N^{O(r \times \log |\Sigma|)})
=O(N^{O(r \times \log |\Sigma|)})$
time.
Other steps take less than that time.
Thus, the total  time  required is
$O(N^{O(r\times \log |\Sigma|)})$,
which is polynomial in term of
input size for any constant $r$.

 From Lemma \ref{KEY}, the performance ratio
of the algorithm is $1+\frac{1}{2r-1}$. $\qed$
\end{proof}

\section{A PTAS For {\sc Closest Substring}}
In this section, we further extend the algorithms
for {\sc Closest String} to a PTAS for {\sc Closest Substring},
making use of a {\em random sampling} strategy.
Note that Algorithm~smallSubstring runs in exponential time
for general radius $d$.  And Algorithm~closestString does not
work for {\sc Closest Substring} since we do not
know how to construct an optimal problem similar to~(\ref{lps1})
--- The construction of~(\ref{lps1}) requires us to know all the $n$
strings (substrings)
in an optimal solution of {\sc Closest String} ({\sc Closest Substring}).
It is easy to see that the choice of a ``good'' substring
from every string $s_i$ is the only obstacle on the way to the solution.
We use random sampling to handle this.

Now let us outline the main ideas.
Let $\langle {\cal S}=\{s_1,s_2,\ldots,s_n\},L \rangle$ be
an instance of
{\sc Closest Substring}, where  $s_i$ is of  length $m$.
Suppose that $s$ is its optimal center string and
$t_i$ is  a length $L$  substring of $s_i$ which is
the closest to $s$ ($i=1,2,\ldots,n$).
Let $d_{opt}=\max _{i=1}^n d(s,t_i)$.
By trying all possibilities, we can assume that
$t_{i_1},t_{i_2},\ldots,t_{i_r}$ are the $r$ substrings $t_{i_j}$
that satisfy Lemma~\ref{KEY} by replacing $s_i$ by $t_i$ and $s_{i_j}$ by $t_{i_j}$.
Let $Q$ be the set of positions where $t_{i_1},t_{i_2},\ldots,t_{i_r}$
agree and $P=\{1,2,\ldots,L\}-Q$.
By Lemma~\ref{KEY}, $t_{i_1}|_Q$ is a good approximation to $s|_Q$.
We want to approximate $s|_P$ by the solution $x$ of the following
optimization problem~(\ref{opt2}), where $t'_i$ is a substring of $s_i$ and
is up to us to choose.
\begin{equation}
\label{opt2}
\left\{
 \begin{array}{l}
 \min \;\; d;\\
 d(t'_i|_P,x) \leq d - d(t'_i|_Q,t_{i_1}|_Q), \;\;
  i=1, \cdots , n; |x|=|P|.
 \end{array}
\right.
\end{equation}

The ideal choice is
$t'_i=t_i$, {\em i.e.}, $t'_i$ is the closest to $s$ among
all substrings of $s_i$.
However, we only approximately know $s$ in $Q$ and
know nothing about $s$ in $P$ so far.
So, we randomly pick $O(\log (mn))$ positions from $P$.
Suppose the multiset of these random positions is $R$.
By trying all possibilities, we can assume that
we know $s$ at these $|R|$ positions.
We then find the substring $t'_i$ from $s$ such that
$d(s|_R,t'_i|_{R})\times \frac {|P|}{|R|}+d(t_{i_1}|_Q,t'_i|_Q)$
is minimized.  Then $t'_i$ potentially belongs to the substrings
which are the closest to $s$.

Then we solve (\ref{opt2}) approximately by the method provided in
the proof of Lemma~\ref{lem-rest} and
combine the solution $x$ at $P$ and $t_{i_1}$ at $Q$, the
resulting string should be a good approximation to $s$.
The detailed algorithm (Algorithm closestSubstring)
is given in Figure~\ref{fig-alg}.
We prove Theorem~\ref{th-ptas} in the rest of the section.

\begin{figure}[ht]
\begin{center}
{\normalfont\normalsize
\begin{tabular}{|l|}
\hline
\multicolumn{1}{|c|}{\bf Algorithm ~closestSubstring}
\\
\makebox[.55in][l]{{Input}} \parbox[t]{4.45in}
{$n$ sequences $\{s_1, s_2,\ldots,s_n\} \subseteq \Sigma^m$, integer $L$.}
\\
\makebox[.55in][l]{{Output}} \parbox[t]{4.45in}
{the center string $s$.}
\\
\makebox[.2in][l]{1.} \parbox[t]{4.8in}
{{\bf for} every $r$ length-$L$ substrings
$t_{i_1}, t_{i_2},\ldots, t_{i_r}$ (allowing repeats, but if $t_{i_j}$ and
$t_{i_k}$ are both chosen from the same $s_i$ then $t_{i_j}=t_{i_k}$)
of $s_1,\ldots ,s_n$
{\bf do}}
\\
\makebox[.5in][r]{(a)} \parbox[t]{4.5in}
{$Q=\{1\leq j \leq L \,|\, t_{i_1}[j]=t_{i_2}[j]=\ldots = t_{i_r}[j] \}$,
$P=\{1,2,\ldots, L\} - Q$.}
\\
\makebox[.5in][r]{(b)} \parbox[t]{4.5in}
{Let $R$ be a multiset containing
$\lceil \frac {4}{\epsilon ^2} \log(nm)\rceil$
uniformly random positions from $P$.
}
\\
\makebox[.5in][r]{(c)} \parbox[t]{4.5in}
{{\bf for} every string $y$ of length $|R|$ {\bf do}}
\\
\makebox[.7in][r]{(i)} \parbox[t]{4.3in}
{{\bf for} $i$ from $1$ to $n$ {\bf do}}
\\
\makebox[.8in][r]{} \parbox[t]{4.2in}
{Let $t'_i$ be a length $L$ substring of $s_i$
minimizing $d(y,t'_i|_{R})\times \frac {|P|}{|R|}+d(t_{i_1}|_Q,t'_i|_Q)$.
}
\\
\makebox[.7in][r]{(ii)} \parbox[t]{4.3in}
{Using the method provided in the proof of
Lemma~\ref{lem-rest}, solve the optimization
problem defined by Formula~(\ref{opt2}) approximately.
Let $x$ be the approximate solution within error $\epsilon\, |P|$.}
\\
\makebox[.7in][r]{(iii)} \parbox[t]{4.3in}
{Let $s'$ be the string such that $s'|_P=x$ and $s'|_Q=t_{i_1}|_Q$.
Let $c=\max^n_{i=1} \min_{\{t_i{\rm ~is~a~substring~of~}s_i\}} d(s',t_i)$.}
\\
\makebox[.2in][l]{2.} \parbox[t]{4.8in}
{{\bf for} every length-$L$ substring $s'$ of $s_1$ {\bf do}}
\\
\makebox[.5in][r]{} \parbox[t]{4.5in}
{Let $c=\max^n_{i=1} \min_{\{t_i{\rm ~is~a~substring~of~}s_i\}} d(s',t_i)$.}
\\
\makebox[.2in][l]{3.} \parbox[t]{4.8in}
{Output the $s'$ with minimum $c$ in step 1(c)(iii) and step 2.}
\\
\hline
\end{tabular}
\caption{The PTAS for the closest substring problem.}
\label{fig-alg}
}
\end{center}
\end{figure}

\begin{theorem}
\label{th-ptas}
Algorithm closestSubstring is a PTAS for the closest substring problem.
\end{theorem}
\begin{proof}
Let $s$ be an optimal center string and $t_i$ be the
length-$L$ substring of $s_i$ that is the closest to $s$.  Let
$d_{opt}=\max d(s,t_i)$.  Let $\epsilon$ be any small
positive number and $r \geq 2$ be any fixed integer.
Let $\rho_0 = \max _{1\leq i,j\leq n} {d(t_i,t_j)}/{d_{opt}}$.
If $\rho _0 \leq 1+\frac{1}{2r-1}$, then clearly we can find a solution
$s'$ within ratio $\rho _0$ in step 2.
So, we assume that $\rho _0 \geq 1+\frac{1}{2r-1}$ from now on.

By Lemma~\ref{KEY}, Algorithm~closestSubstring picks
a group of $t_{i_1},t_{i_2},\ldots,t_{i_r}$ in step 1
at some point such that

\vspace{1ex}
\noindent
{\bf Fact 1~}
For any $1 \leq l \leq n$,
$
|\{j \in Q\,|\,
 t_{i_1}[j] \neq t_l[j] \mbox{ and } t_{i_1}[j]
\neq s[j]\} |
\leq  \frac{1}{2r-1} \,d_{opt}.
$
\vspace{1ex}

Obviously, the algorithm takes $y$ as $s|_R$ for at some point
in step 1(c).  Let $y=s|_R$ and $t_{i_1},t_{i_2},\ldots,t_{i_r}$
satisfy Fact~1.  Let $t'_i$ be defined as in step 1(c)(i).
Let $s^*$ be a string such that $s^*|_P=s|_P$ and
$s^*|_Q=t_{i_1}|_Q$.  Then we claim:

\vspace{1ex}
\noindent
{\bf Fact 2~}
With high probability,
$d(s^*,t'_i)\leq d(s^*,t_i)+ 2 \epsilon |P|$
for all $1\leq i\leq n$.\\

\begin{proof}
For convenience, for any position multiset $T$,
we denote $d^T(t_1,t_2)=d(t_1|_T,t_2|_T)$ for any two
strings $t_1$ and $t_2$.  Let $\rho=\frac{|P|}{|R|}$.
Consider  any length $L$ substring $t'$ of $s_i$
satisfying
\begin{equation}
d(s^*, t')\geq d(s^*,t_i)+2 \epsilon |P|.
\label{eq-close150}
\end{equation}
It is easy to see that
$
 \rho\, d^R(s^*,t') + d^Q(t_{i_1},t')
\leq \rho\, d^R(s^*,t_i)+ d^Q(t_{i_1},t_i)$ implies
either
$(\rho \, d^R(s^*,t') + d^Q(s^*,t') \leq d(s^*,t') -\epsilon |P|$
or
$\rho \, d^R(s^*,t_i)+ d^Q(s^*,t_i) \geq d(s^*,t_i) + \epsilon |P|$.
Thus, we have the following inequality:
\begin{eqnarray}
&&{\bf Pr} \left( \rho\, d^R(s^*,t') + d^Q(t_{i_1},t')
\leq \rho\, d^R(s^*,t_i)+ d^Q(t_{i_1},t_i) \right)
\nonumber\\
&\leq& {\bf Pr}\left(\rho \, d^R(s^*,t') + d^Q(s^*,t')
\leq d(s^*,t') -\epsilon |P|\right)+
\nonumber\\
&&{\bf Pr}\left(\rho \, d^R(s^*,t_i)+ d^Q(s^*,t_i)
\geq d(s^*,t_i) + \epsilon |P|\right).
\label{eq-close200}
\end{eqnarray}

It is easy to see that $d^R(s^*,t')$ is the  sum of $|R|$ independent
random 0-1 variables $ \sum _{i=1} ^{|R|} X_i$,  where
$X_i=1$ indicates  a mismatch between  $s^*$ and  $t'$ at
the $i$-th position in $R$.
Let $\mu = E[d^R(s^*,t')]$.
Obviously, $\mu=d^P(s^*,t') / \rho$.
Therefore, by Lemma~\ref{lem-chernoff1} (2),
\begin{eqnarray}
&&{\bf Pr}\left(\rho \, d^R(s^*,t') + d^Q(s^*,t') \leq d(s^*,t') -\epsilon |P|\right)
\nonumber\\
&=&{\bf Pr}\left(d^R(s^*,t') \leq (d(s^*,t')- d^Q(s^*,t'))/\rho -\epsilon |R|\right)
\nonumber\\
&=&{\bf Pr}\left(d^R(s^*,t') \leq d^P(s^*,t')/\rho -\epsilon |R|\right)
\nonumber\\
&=&{\bf Pr}\left(d^R(s^*,t') \leq \mu -\epsilon |R|\right)
\leq \exp\left(-\frac 12 \epsilon ^2 |R| \right) \leq (nm)^{-2},
\label{eq-close300}
\end{eqnarray}
where the last inequality is
due to the setting
$|R|=\lceil \frac {4}{\epsilon ^2}\log (nm)\rceil$ in
step 1(b) of the algorithm.
Similarly,  using  Lemma~\ref{lem-chernoff1} (1)  we have
\begin{equation}
\label{eq-close400}
{\bf Pr}\left(\rho \, d^R(s^*,t_i)+ d^Q(s^*,t_i)
\geq d(s^*,t_i) + \epsilon |P|\right)
\leq (nm)^{-\frac 43}.
\end{equation}
Combining Formula~(\ref{eq-close200})(\ref{eq-close300})(\ref{eq-close400}),
we know that for any $t'$ that satisfies Formula~(\ref{eq-close150}),
\begin{equation}
{\bf Pr} \left( \rho\, d^R(s^*,t') + d^Q(t_{i_1},t')
\leq \rho\, d^R(s^*,t_i)+ d^Q(t_{i_1},t_i) \right)
\leq 2\, (nm)^{-\frac 43}.
\label{eq-close500}
\end{equation}
For any fixed $1\leq i\leq n$, there are less than $m$ substrings $t'$
that satisfies Formula~(\ref{eq-close150}).  Thus,
from Formula~(\ref{eq-close500}) and the definition of $t'_i$,
\begin{equation}
{\bf Pr}\left( d(s^*,t'_i) \geq d(s^*,t_i)+ 2 \epsilon |P| \right)
\leq 2\,n^{-\frac {4}{3}}m^{-\frac 13}.
\end{equation}
Summing up all $i\in [1,n]$, we know that with probability
at least $1-2\,(nm)^{-\frac 13}$,
$d(s^*,t'_i)\leq d(s^*,t_i)+ 2 \epsilon |P|$ for all $i$.
$\qed$
\end{proof}

 From Fact 1,
$d(s^*,t_i)=d^P(s,t_i)+d^Q(t_{i_1},t_i)\leq d(s,t_i)+\frac 1{2r-1} \, d_{opt}.$
Combining with Fact~2 and $|P|\leq r\, d_{opt}$,
we get
\begin{equation}
d(s^*,t'_i)\leq (1+ \frac 1{2r-1} + 2 \epsilon \, r) d_{opt}.
\label{eq-close600}
\end{equation}
By the definition of $s^*$, the optimization
problem defined by Formula~(\ref{opt2}) has a solution $s|_P$
such that $d\leq (1+ \frac 1{2r-1} + 2 \epsilon \, r) d_{opt}$.
We can solve  the optimization problem  within error $\epsilon |P|$ by the method in
the proof of Lemma~\ref{lem-rest}.
Let  $x$ be  the solution of the optimization problem.
Then by Formula~(\ref{opt2}), for any $1\leq i\leq n$,
\begin{equation}
d(t'_i|_P,x)\leq
(1+ \frac 1{2r-1} + 2 \epsilon \, r) d_{opt}
-d(t'_i|_Q,t_{i_1}|_Q)+\epsilon |P|.
\label{close-800}
\end{equation}
Let $s'$ be defined in step 1(c)(iii), then by Formula~(\ref{close-800}),
\begin{eqnarray*}
d(s',t'_i)&=&d(x,t'_i|_P)+ d(t_{i_1}|_Q,t'_i|_Q)
\\&\leq& (1+ \frac 1{2r-1} + 2 \epsilon r) d_{opt} + \epsilon |P|
\\&\leq& (1+\frac 1{2r-1} + 3\epsilon r) d_{opt}.
\end{eqnarray*}

It is easy to see that the algorithm runs in polynomial time for
any fixed positive $r$ and $\epsilon$.  For any $\delta>0$, by
properly setting $r$ and $\epsilon$ such that
$\frac 1{2r-1} + 3\epsilon r \leq \delta$, with high probability,
the algorithm outputs in polynomial time a solution $s'$
such that $d(t'_i,s')$ is no more than $(1+\delta)d_{opt}$
for every $1\leq i\leq n$, where $t'_i$ is a substring of $s_i$.
The algorithm can be derandomized by standard methods~\cite{MR95}.
$\qed$
\end{proof}

\section*{Acknowledgements}
We would like to thank Tao Jiang, Kevin Lanctot,
Joe Wang, and Louxin Zhang for discussions and suggestions on related
topics.

Ming Li is supported in part by the
NSERC Research Grant OGP0046506, a CGAT grant,
the E.W.R. Steacie Fellowship. Bin Ma is supported in part by
the NSERC Research Grant OGP0046506. Bin Ma and Lusheng Wang are
supported in part by HK RGC Grants 9040297, 9040352, 9040444 and CityU
Strategic Grant 7000693.

\end{document}